%% file: Hubbard-PRA-sub2.tex
\documentclass[pra,
   reprint,
   superscriptaddress,
   showpacs,
   preprintnumbers,
   ]{revtex4-1}
\input{Johnsdefs}
\begin{document}
%
%
%
%
\title[LOAF Bose-Hubbard model]
   {Leading-Order Auxiliary Field Theory of the Bose-Hubbard Model}
\author{John F. Dawson}
\email{john.dawson@unh.edu}
\affiliation{Department of Physics,
   University of New Hampshire,
   Durham, NH 03824}
\author{Fred Cooper} 
\email{fcooper@fas.harvard.edu}
\affiliation{Department of Earth and Planetary Science, Harvard University,Cambridge, MA 02138}
\affiliation{The Santa Fe Institute, Santa Fe, NM 87501, USA}
\author{Chih-Chun Chien}
\email{chihchun@lanl.gov}
\affiliation{
   Los Alamos National Laboratory,
   Los Alamos, NM 87545}
\author{Bogdan Mihaila}
\email{bmihaila@nsf.gov}
\affiliation{
   Los Alamos National Laboratory,
   Los Alamos, NM 87545}
\affiliation{National Science Foundation,
                Arlington, VA 22230}

\date{\today, \now \ EST}


\begin{abstract}
We discuss the phase diagram of the Bose-Hubbard (BH) model in the leading-order auxiliary field (LOAF) theory.  LOAF is a conserving non-perturbative approximation that treats on equal footing the normal and anomalous density condensates. The mean-field solutions in LOAF correspond to first-order and second-order phase transition solutions with two critical temperatures corresponding to a vanishing Bose-Einstein condensate, $T_c$, and a vanishing diatom condensate, $T^\star$. 
The \emph{second-order} phase transition solution predicts the correct order of the transition in continuum Bose gases. 
For either solution, the superfluid state is tied to the presence of the diatom condensate related to the anomalous density in the system.  
In ultracold Bose atomic gases confined on a three-dimensional lattice, the critical temperature $T_c$ exhibits a quantum phase transition, where $T_c$ goes to zero at a finite coupling. The BH phase diagram in LOAF features a line of first-order transitions ending in a critical point beyond which the transition is second order while approaching the quantum phase transition. We identify a region where a diatom condensate is expected for temperatures higher than $T_c$ and less than $T_0$, the critical temperature of the non-interacting system.
The LOAF phase diagram for the BH model 
compares qualitatively well with existing experimental data and results of \emph{ab initio} Monte Carlo simulations. 
\end{abstract}
\maketitle
%
%
\section{\label{s:intro}Introduction}

The Bose-Hubbard (BH) model has been the subject of broad theoretical~\cite{PhysRevB.40.546,PhysRevB.53.2691,PhysRevA.68.043623,PhysRevA.71.033608,PhysRevB.75.134302,PhysRevB.80.245110,PhysRevA.85.063602} and experimental~\cite{PhysRevLett.86.4447,Greiner02,r:Trotzky:2010fk} interest. In addition to being a challenge for many-body theories~\cite{RevModPhys.80.885}, its realization in ultracold atoms~\cite{Greiner02} opened opportunities for studying the BH model in a controllable and accurate way. Developing and improving mean-field descriptions for the BH model has been an important task since Fischer \emph{et al.}~\cite{PhysRevB.40.546} discussed the zero-temperature mean-field phase diagram for the BH model. Studies of the BH model have been summarized in many textbooks~\cite{PethickBEC,Leggettbook,Uedabook}. One may also test various many-body theoretical techniques using the BH model and benchmark those methods. Here we follow this tradition and study the BH model using a well-developed theoretical framework.

Recently, we introduced a leading-order auxiliary field (LOAF) theory for a homogeneous system of ultracold gas of bosonic atoms~\cite{PhysRevLett.105.240402,PhysRevA.83.053622}. To derive this formalism, we used the Hubbard-Stratonovitch transformation~\cite{r:Hubbard:1959kx,r:Stratonovich:1958vn} to introduce auxiliary fields related to the normal and anomalous density condensates. Path integral methods were used to obtain a leading-order expansion of the partition function using the auxiliary fields to organize the expansion method. The resulting non-perturbative mean-field theory produces a conserving and gapless approximation that is applicable to large interval of coupling-constant values, satisfies the Goldstone theorem, yields a Bose-Einstein transition that is second order, and predicts a positive shift in the critical temperature, $T_c$, consistent with other similar methods~\cite{r:Baym:2000fk}. The relation of the LOAF theory to the Goldstone theorem and the Higgs mechanism was discussed in Ref.~\onlinecite{PhysRevA.85.023631}.  The behavior of the LOAF theory near the critical point is discussed in Ref.~\onlinecite{PhysRevA.84.023603} and the relation to superfluidity and the Josephson relation is discussed in Ref.~\onlinecite{PhysRevA.86.013603}, where we showed that the superfluid density in LOAF is proportional with the square of the anomalous-density \emph{diatom} condensate. The latter is analogous with the Cooper-pair condensate in the BCS mean-field theory of dilute Fermi gases~\cite{PhysRevB.77.144521}.

In this paper we develop the LOAF theory of the Bose-Hubbard model, which has been used to study the physics of ultracold Bose atoms in optical lattices~\cite{PhysRevA.71.033608,PhysRevB.75.134302,PhysRevB.80.245110,PhysRevA.85.063602}. Perhaps the most salient feature of the BH model is the prediction of a superfluid to Mott insulator phase transition. The latter was demonstrated experimentally in three-dimensional optical lattices by Trotzky \emph{et al.}~\cite{r:Trotzky:2010fk} by observing the suppression of the critical temperature for superfluidity near the Mott transition. These experimental results were showed to compare nicely with theoretical predictions based on quantum Monte Carlo simulations~\cite{PhysRevB.75.134302}. Monte Carlo also predicted the critical interaction strength for which the critical temperature goes to zero and the phase transition is purely quantum in character~\cite{PhysRevB.75.134302}. 

We will show that below the critical temperature, $T_c$, where the \emph{usual} Bose-Einstein condensate vanishes, LOAF is identical with the Hamiltonian version of LOAF introduced recently by Kleinert, Narzikulov, and Rakhimov~\cite{r:Kleinert:2013fk} and referred to as the ``two-collective'' field  theory for the BH model. Above $T_c$, LOAF has two possible solutions corresponding to first-order and second-order phase transitions, respectively. 
For both solutions, the superfluid state is indicated by the presence of an anomalous-density \emph{diatom} condensate in the system. 
In ultracold Bose atomic gases confined on a three-dimensional lattice, LOAF predicts that the critical temperature $T_c$ exhibits a quantum phase transition (QPT), where $T_c$ goes to zero at a finite coupling. The BH phase diagram in LOAF features a line of first-order transitions ending in a critical point beyond which the transition is second order while approaching the QPT limit. We identify a region where a diatom condensate is expected for temperatures $T_c < T < T_0$. Here, $T_0$ is the critical temperature of the non-interacting system. Overall, the LOAF phase diagram is very similar to the superfluid to Mott insulator transition in systems of ultracold Bose atoms trapped in optical lattices and compares qualitatively well with results of \emph{ab initio} Monte Carlo simulations~\cite{PhysRevB.75.134302} and available experimental data~\cite{r:Trotzky:2010fk}.  We analyze numerically the properties of the LOAF theory in the weak-coupling limit.

%
%
\section{\label{s:Hubbard}The Bose-Hubbard model}

We consider the case of $N$ bosonic atoms trapped in a three-dimensional cubic lattice. 
%
%
\subsection{\label{ss:RealTime}Real time formulation}

The path integral for the boson field $\phi(\bx,t)$ is given by
\begin{gather}
   Z[ \, j,j^{\ast} \, ]
   =
   e^{i W[ \, j,j^{\ast} \, ] / \hbar}
   =
   \iint \rD \phi \, \rD \phi^{\ast} \,
   e^{i S[\, \phi,\phi^{\ast};j,j^{\ast} \, ] / \hbar} \>,
   \notag \\
   S[\, \phi,\phi^{\ast};j,j^{\ast} \, ]
   =
   \int \! \rd t \> L[ \, \phi,\phi^{\ast};j,j^{\ast} \, ] \>,
   \label{BH.e:eq1}
\end{gather}
where the Lagrangian is
\begin{align}
   &L[ \, \phi,\phi^{\ast};j,j^{\ast} \, ]
   \label{BH.e:eq2} \\
   &=
   \frac{i \hbar}{2} \, 
   \int \rd^3 x \,
   \bigl \{ \,
      \phi^{\ast}(\bx,t) \, [ \partial_{t} \phi(\bx,t) ]
      -
      [ \partial_{t} \phi^{\ast}(\bx,t) ] \, \phi(\bx,t) \,
   \bigr \}
   \notag \\
   & \quad
   -
   \int \rd^3 x \,
   \phi^{\ast}(\bx,t) \,
   \Bigl [
      -
      \frac{\hbar^2 \nabla^2}{2m}
      +
      V(\bx) \,
   \Bigr ]
   \phi(\bx,t)
   \notag \\
   &
   -
   \frac{1}{2}
   \iint \rd^3 x \, \rd^3 x' \,
   \phi^{\ast}(\bx,t) \, \phi^{\ast}(\bx',t) \, 
   U(\bx,\bx') \, 
   \phi(\bx',t) \, \phi(\bx,t)
   \notag
   \notag \\
   & \qquad
   +
   \int \rd^3 x \,
   \bigl [ \,
      j^{\ast}(\bx,t) \, \phi(\bx,t)
      +
      j(\bx,t) \, \phi^{\ast}(\bx,t) \,
   \bigr ] \>.
   \notag
\end{align}
Variation of the action yields a Schr{\"o}dinger equation for the field $\phi(\bx,t)$.
Here $V(\bx)$ is the parodic potential created by the optical lattice and $U(\bx,\bx')$ is the interaction energy between atoms.  The tight-binding approximation assumes that the field $\phi(\bx,t)$ can be expanded in normalized eigenfunctions $\psi(\bx)$ of an atom trapped in the lattice at positions $\bx_{\bi} = a \, \bi$, 
\begin{equation}\label{BH.e:eq3}
   \phi(\bx,t)
   =
   \sum_{\bi} \,
   \phi_{\bi}(t) \, \psi(\bx - \bx_{\bi}) \>,
\end{equation}
where $\psi(\bx)$ satisfies
\begin{equation}\label{BH.e:eq4}
   \Bigl [ \,
      - 
      \frac{\hbar^2 \nabla^2}{2m}
      +
      V(\bx) \,
   \Bigr ] \,
   \psi(\bx)
   =
   E \, \psi(\bx) \>,
\end{equation}
where $E$ is the ground state energy of a trapped atom of mass $m$.
Here $\bi = ( i_x,i_y,i_z )$ are triplets of integers, each running from $1$ to $N_s$.  Inversion of \eqref{BH.e:eq3} gives
\begin{equation}\label{BH.e:eq5}
   \phi_{\bi}(t)
   =
   \int \rd^3 x \,
   \psi^{\ast}(\bx - \bx_{\bi}) \, \phi(\bx,t) \>.
\end{equation}
So using expansion \eqref{BH.e:eq3}, and keeping overlaps with nearest neighbors, we find
\begin{align}
   &\int \rd^3 x \,
   \phi^{\ast}(\bx,t) \,
   \Bigl [
      -
      \frac{\hbar^2 \nabla^2}{2m}
      +
      V(\bx) \,
   \Bigr ]
   \phi(\bx,t)
   \label{BH.e:eq6} \\
   & \qquad
   \approx
   \sum_{\bi} \, 
   \Bigl \{ \,
      E \, | \phi_{\bi}(t) |^2
      \notag \\
      & \qquad\qquad
      -
      J
      \sum_{\kappa}
      \bigl [ \,
         \phi_{\bi}^{\ast}(t) \, \phi_{\bi + \kappa}^{\phantom\ast}(t)
         +
         \phi_{\bi+\kappa}^{\ast}(t) \, \phi_{\bi}^{\phantom\ast}(t) \,
      \bigr ] \,
   \Bigr \} \>,
   \notag   
\end{align}
where $\kappa = (1,0,0),(0,1,0),(0,0,1)$ is the displacement by one unit in the $(x,y,z)$-directions, and $J$ is the overlap integral,
\begin{equation}\label{BH.e:eq7}
   J
   =
   -
   \int \rd^3 x \,
   \psi(\bx + a) \, V(\bx) \, \psi(\bx) > 0 \>.
\end{equation}
The particle-particle interaction is assumed to be a short range contact interaction,
\begin{align}
   &\frac{1}{2}
   \iint \rd^3 x \, \rd^3 x' \,
   \phi^{\ast}(\bx,t) \, \phi^{\ast}(\bx',t) \, 
   U(\bx,\bx') \, 
   \phi(\bx',t) \, \phi(\bx,t)
   \notag \\
   & \qquad\qquad
   \approx
   \frac{U}{2} \,
   \sum_{\bi} \,
   | \phi_{\bi}(t) \, |^4 \>.
   \label{BH.e:eq7.1} 
\end{align}
So in the tight-binding approximation, the Lagrangian is given by
\begin{align}
   &L[ \, \phi^{\phantom\ast}_{\bi},\phi_{\bi}^{\ast};
          j^{\phantom\ast}_{\bi},j_{\bi}^{\ast} \, ]
   \label{BH.e:eq8} \\
   &
   =
   \frac{i \hbar}{2} \, 
   \sum_{\bi} \,
   \bigl \{ \,
      \phi^{\ast}_{\bi}(t) \, 
      [ \partial^{\phantom\ast}_{t} \phi^{\phantom\ast}_{\bi}(t) ]
      -
      [ \partial^{\phantom\ast}_{t} \phi^{\ast}_{\bi}(t) ] \, 
      \phi^{\phantom\ast}_{\bi}(t) \,
   \bigr \}
   \notag \\
   &
   - \!\!
   \sum_{\bi}
   \Bigl \{
      E \, | \phi_{\bi}(t) |^2
      -
      J
      \sum_{\kappa}
      \bigl [ 
         \phi_{\bi}^{\ast}(t) \, \phi_{\bi + \kappa}^{\phantom\ast}(t)
         +
         \phi_{\bi+\kappa}^{\ast}(t) \, \phi_{\bi}^{\phantom\ast}(t)
      \bigr ]
   \Bigr \}
   \notag \\
   & \qquad
   -
   \frac{U}{2} \,  
   \sum_{\bi} \, | \phi_{\bi}(t) \, |^4 
   +
   \sum_{\bi} \, 
   \bigl [ \,
      j^{\ast}_{\bi}(t) \, \phi^{\phantom\ast}_{\bi}(t)
      +
      j^{\phantom\ast}_{\bi}(t) \, \phi^{\ast}_{\bi}(t) \,
   \bigr ] \>,
   \notag
\end{align}
which we can write as:
\begin{align}
   &L[ \, \phi^{\phantom\ast}_{\bi},\phi_{\bi}^{\ast};
          j^{\phantom\ast}_{\bi},j_{\bi}^{\ast} \, ]
   \label{BH.e:eq9} \\
   &=
   \frac{i \hbar}{2} \, 
   \sum_{\bi} \,
   \bigl \{ \,
      \phi^{\ast}_{\bi}(t) \, 
      [ \partial^{\phantom\ast}_{t} \phi^{\phantom\ast}_{\bi}(t) ]
      -
      [ \partial^{\phantom\ast}_{t} \phi^{\ast}_{\bi}(t) ] \, 
      \phi^{\phantom\ast}_{\bi}(t) \,
   \bigr \}
   \notag \\
   & \qquad
   -
   \sum_{\bi} \, ( \, E - 2d \, J \, ) \, | \phi_{\bi}(t) |^2
   -
   \frac{U}{2} \,  
   \sum_{\bi} \, | \phi_{\bi}(t) \, |^4 
   \notag \\
   &
   -
   J
   \sum_{\bi,\kappa} \,
   \Bigl \{ \,
      2 \, | \phi_{\bi}(t) |^2
      -
      \bigl [ \,
         \phi_{\bi}^{\ast}(t) \, \phi_{\bi + \kappa}^{\phantom\ast}(t)
         +
         \phi_{\bi+\kappa}^{\ast}(t) \, \phi_{\bi}^{\phantom\ast}(t) \,
      \bigr ] \,
   \Bigr \}
   \notag \\
   & \qquad
   +
   \sum_{\bi} \, 
   \bigl [ \,
      j^{\ast}_{\bi}(t) \, \phi^{\phantom\ast}_{\bi}(t)
      +
      j^{\phantom\ast}_{\bi}(t) \, \phi^{\ast}_{\bi}(t) \,
   \bigr ] \>,
   \notag
\end{align}
Here $d$ is the number of spatial dimensions.
The constant energy term proportional to $E - 2d J$ can be eliminated by changing variables to
\begin{equation}\label{BH.e:eq10}
   \phi_{\bi}(t) 
   =
   e^{- i ( E - 2d J ) \, t / \hbar } \, \tilde{\phi}_{\bi}(t) \>,
\end{equation}
which simply changes the energy scale.
Then in terms of these new variables, \eqref{BH.e:eq9} becomes
\begin{align}
   &L[ \, \tilde{\phi}^{\phantom\ast}_{\bi},\tilde{\phi}_{\bi}^{\ast};
          \tilde{j}^{\phantom\ast}_{\bi},\tilde{j}_{\bi}^{\ast} \, ]
   \label{BH.e:eq11} \\
   &=
   \frac{i \hbar}{2} \, 
   \sum_{\bi} \,
   \bigl \{ \,
      \tilde{\phi}^{\ast}_{\bi}(t) \, 
      [ \partial^{\phantom\ast}_{t} \tilde{\phi}^{\phantom\ast}_{\bi}(t) ]
      -
      [ \partial^{\phantom\ast}_{t} \tilde{\phi}^{\ast}_{\bi}(t) ] \, 
      \tilde{\phi}^{\phantom\ast}_{\bi}(t) \,
   \bigr \}
   \notag \\
   &
   -
   J
   \sum_{\bi,\kappa} \, 
   \Bigl \{ \,
      2 \, | \tilde{\phi}_{\bi}(t) |^2
      -
      \bigl [ \,
         \tilde{\phi}_{\bi}^{\ast}(t) \, \tilde{\phi}_{\bi + \kappa}^{\phantom\ast}(t)
         +
         \tilde{\phi}_{\bi+\kappa}^{\ast}(t) \, \tilde{\phi}_{\bi}^{\phantom\ast}(t) \,
      \bigr ] \,
   \Bigr \}
   \notag \\
   &
   -
   \frac{U}{2} \,  
   \sum_{\bi} \, | \tilde{\phi}_{\bi}(t) \, |^4 
   +
   \sum_{\bi} \, 
   \bigl [ \,
      \tilde{j}^{\ast}_{\bi}(t) \, \tilde{\phi}^{\phantom\ast}_{\bi}(t)
      +
      \tilde{j}^{\phantom\ast}_{\bi}(t) \, \tilde{\phi}^{\ast}_{\bi}(t) \,
   \bigr ] \>.
   \notag
\end{align}
Eq.~\eqref{BH.e:eq11} is the usual form of the Bose-Hubbard Lagrangian.  From now on we drop the tilde notation.

%
%
\subsection{\label{ss:ImagTime}Imaginary time formulation}

The imaginary time action is obtained from the real time action by the mapping $t \mapsto - i \hbar \tau$ and $L \mapsto - L_{\text{E}}$.  The partition function $Z$ for the BH model is then written as
\begin{gather}
   Z[ \, j,j^{\ast} \, ]
   =
   e^{ - \beta \, \Omega[ \, j,j^{\ast} \, ] }
   =
   \iint \rD \phi \, \rD \phi^{\ast} \,
   e^{ - S_{\text{E}}[\, \phi,\phi^{\ast};j,j^{\ast} \, ] } \>,
   \notag \\
   S_{\text{E}}[\, \phi,\phi^{\ast};j,j^{\ast} \, ]
   =
   \int_0^{\beta} \!\!\! \rd \tau \> 
   L_{\text{E}}[ \, \phi,\phi^{\ast};j,j^{\ast} \, ] \>,
   \label{BH.e:eq12}
\end{gather}
where the Euclidean Lagrangian is
\begin{align}
   &L_{\text{E}}[ \, \phi,\phi^{\ast};j,j^{\ast} \, ]
   \label{BH.e:eq13} \\
   &=
   \sum_{\bi} \,
   \Bigl \{ \,
   \frac{1}{2} \, 
   \bigl \{ \,
      \phi_{\bi}^{\ast}(\tau) \, [ \partial_{\tau} \phi_{\bi}(\tau) ]
      -
      [ \partial_{\tau} \phi_{\bi}^{\ast}(\tau) ] \, \phi_{\bi}(\tau) \,
   \bigr \}
   \notag \\
   &
   +
   J
   \sum_{\kappa} \, 
   \bigl \{ \,
      2 \, | \phi_{\bi}(t) |^2
      -
      \bigl [ \,
         \phi_{\bi}^{\ast}(\tau) \, \phi_{\bi + \kappa}^{\phantom\ast}(\tau)
         +
         \phi_{\bi+\kappa}^{\ast}(\tau) \, \phi_{\bi}^{\phantom\ast}(\tau) \,
      \bigr ] \,
   \bigr \}
   \notag \\
   &
   +
   \frac{U}{2} \, | \phi_{\bi}(\tau) \, |^4 
   -
   \mu \, | \phi_{\bi}(\tau) |^2 
   -
   j^{\ast}_{\bi}(\tau) \, \phi^{\phantom\ast}_{\bi}(\tau)
   -
   j^{\phantom\ast}_{\bi}(\tau) \, \phi^{\ast}_{\bi}(\tau) \,
   \Bigl \} \>.
   \notag
\end{align}
Here we have dropped the tilde notation and introduced a chemical potential $\mu$.

%
%
\section{\label{sss:Auxfields}LOAF formalism}

In the leading-order auxiliary field (LOAF) method, we introduce two auxiliary fields $\chi_{\bi}(\tau)$ and $\Delta_{\bi}(\tau)$ by means of the Hubbard-Stratonovitch transformation~\cite{r:Hubbard:1959kx,r:Stratonovich:1958vn}.  In our case, the auxiliary-field Lagrangian density takes the form
\begin{align}
   L_{\text{aux}}[ \, \Phi,\Delta \, ]
   &=
   \sum_{\bi} \,
   \Bigl \{ \,
   \frac{1}{2 U} \,
   \bigl | \,
      A_{\bi}(\tau) 
      - 
      U \, \phi^{2}_{\bi}(\tau) \,
   \bigr |^2
   \label{BH.e:eq14} \\
   & \qquad
   -
   \frac{1}{2 U} \,
   \bigl [ \,
      \chi_{\bi}(\tau) - U \, \sqrt{2} \, | \phi_{\bi}(\tau) |^2 \,
   \bigr ]^2 \,
   \Bigr \} \>,
   \notag
\end{align}
which we add to Eq.~\eqref{BH.e:eq13}.  We show in Ref.~\onlinecite{PhysRevA.83.053622} that this choice, in the weak coupling limit, agrees with Bogoliubov theory~\cite{r:Bogoliubov:1947ys,r:Andersen:2004uq}.
The action is then becomes
\begin{align}
   &S_{\text{E}}[\Phi,\Delta;J,K] 
   \label{BH.e:eq15} \\
   &=
   \frac{1}{2} \, 
   \int_0^{\beta} \!\!\! \rd \tau \int_0^{\beta} \!\!\! \rd \tau' \sum_{\bi,\bj} \> 
   \Phi^{\dagger}_{\bi}(\tau) \, \calG^{-1}_{\bi,\bj}(\tau,\tau') \, \Phi_{\bj}(\tau')
   \notag \\
   &
   -
   \int_0^{\beta} \!\!\! \rd \tau \sum_{\bi} \, 
   \Bigl \{ \,
      \frac{\chi^2_{\bi}(\tau) - |A_{\bi}(\tau)|^2}{2 U} \,
      +
      J^{\dagger}_{\bi}(\tau) \, \Phi^{\phantom\dagger}_{\bi}(\tau)
      \notag \\
      & \qquad\qquad
      +
      K^{\dagger}_{\bi}(\tau) \, \Delta_{\bi}^{\phantom\dagger}(\tau) \,
   \Bigr \} \>,
   \notag
\end{align}
with
\begin{align}
   &\calG^{-1}_{\bi,\bj}(\tau,\tau')
   \label{BH.e:eq16} \\
   & \qquad
   = 
   \delta(\tau,\tau') \, 
   \begin{pmatrix}
      h_{\bi,\bj}
      + 
      \delta_{\bi,\bj} \, \partial_{\tau} \,,
      &
      - \delta_{\bi,\bj} \, A_{\bi}(\tau) 
      \\[3pt]
      - \delta_{\bi,\bj} \, A^{\ast}_{\bi}(\tau) \,,
      & 
      h_{\bi,\bj} 
      - 
      \delta_{\bi,\bj} \, \partial_{\tau}
   \end{pmatrix} \>,
   \notag
\end{align}
where
\begin{subequations}\label{BH.e:eq16.1}
\begin{align}
   h_{\bi,\bj}
   &=
   J \, \nabla_{\bi,\bj}
   +
   \delta_{\bi,\bj} \, 
   [ \,
      \sqrt{2} \, \chi(\tau) 
      - 
      \mu \,
   ]  \>,
   \label{BH.e:eq16.1a} \\
   \nabla_{\bi,\bj}
   &=
   \sum_{\kappa}
   \bigl \{ \,
      2 \, \delta_{\bi,\bj}
      -
      \bigl [ \,
         \delta_{\bi,\bj+\kappa}
         +
         \delta_{\bi+\kappa,\bj} \,
      \bigr ] \,
   \bigr \} \>.
   \label{BH.e:eq16.1b}
\end{align}
\end{subequations}
Here we have introduced currents which we write as $J_{\bi}(\tau)$ and $K_{\bi}(\tau)$ and a notation,
\begin{equation}\label{BH.e:eq17}
   \Phi_{\bi}(\tau)
   =
   \begin{pmatrix}
      \phi_{\bi}(\tau) \\
      \phi^{\ast}_{\bi}(\tau)
   \end{pmatrix} \>,
   \qquad
   J_{\bi}(\tau)
   =
   \begin{pmatrix}
      j_{\bi}(\tau) \\
      j^{\ast}_{\bi}(\tau)
   \end{pmatrix}
\end{equation}
for the particle fields and currents, and a notation
\begin{equation}\label{BH.e:eq18}
   \Delta_{\bi}(\tau)
   =
   \begin{pmatrix}
      A_{\bi}(\tau) \\ \chi_{\bi}(\tau) \\ A^{\ast}_{\bi}(\tau)
   \end{pmatrix} \>,
   \qquad
   K_{\bi}(\tau)
   =
   \begin{pmatrix}
      k_{\bi}(\tau) \\ k_{0\,\bi}(\tau) \\ k^{\ast}_{\bi}(\tau)
   \end{pmatrix} \>,
\end{equation}
for the auxiliary fields and currents.  
The generating functional for the fields is written as a path integral over all the fields
\begin{equation}\label{BH.e:eq19}
   Z[J,K]
   =
   e^{ - \beta \Omega[J,K]}
   =
   \iint \rD \Phi \, \rD \Delta \,
   e^{ - S_{\text{E}}[\, \Phi,\Delta;J,K \, ] } \>.
\end{equation}
The action is now quadratic in the $\phi_{\bi}(\tau)$ which can be integrated out, giving
\begin{equation}\label{BH.e:eq20}
   Z[J,K]
   =
   \int \rD \Delta \,
   e^{ - S_{\text{eff}}[\, \Delta;J,K \, ] } \>,
\end{equation}
where now
\begin{align}
   &S_{\text{eff}}[\, \Delta;J,K \, ]
   \label{BH.e:eq20.1} \\
   &=
   -
   \frac{1}{2}
   \iint_0^{\beta} \!\!\! \rd \tau \, \rd \tau' \sum_{\bi,\bj} \> 
   J^{\dagger}_{\bi}(\tau) \, \calG_{\bi,\bj}(\tau,\tau') \, J_{\bj}(\tau')
   \notag \\
   & \quad
   -
   \int_0^{\beta} \!\!\! \rd \tau \sum_{\bi} \, 
   \Bigl \{ \,
      \frac{\chi^2_{\bi}(\tau) - |A_{\bi}(\tau)|^2}{2 U}
      +
      K^{\dagger}_{\bi}(\tau) \, \Delta^{\phantom\dagger}_{\bi}(\tau)
      \notag \\
      & \qquad\qquad
      -
      \frac{1}{2} \,
      \Tr{ \Ln{ \calG_{\bi,\bi}^{-1}(\tau,\tau) } } \,
   \Bigr \} \>.
   \notag
\end{align}
Expanding the effective action,
\begin{align}
   &S_{\text{eff}}[\,  \Delta;J,K \, ]
   =
   S_{\text{eff}}[\, \bar{\Delta};J,K \, ]
   \label{BH.e:eq21} \\
   & \qquad
   +
   \int_0^{\beta} \!\!\! \rd \tau \sum_{\bi} \,
   \Bigl [ \,
      \frac{ \delta S_{\text{eff}}[\, \Delta;J,K \, ] }
           { \delta \Delta_{\bi}(\tau) } \,
   \Bigr ]_{\bar{\Delta}}
   ( \, \Delta_{\bi}(\tau) - \bar{\Delta}_{\bi}(\tau) \, )
   \notag \\
   & \qquad
   +
   \frac{1}{2} 
   \iint_0^{\beta} \!\!\! \rd \tau \, \rd \tau' \sum_{\bi,\bj}\,
   \Bigl [ \,
      \frac{ \delta^2 S_{\text{eff}}[\, \Delta;J,K \, ] }
           { \delta \Delta_{\bi}(x) \, \delta \Delta_{\bj}(x') } \,
   \Bigr ]_{\bar{\Delta}}
   \notag \\
   & \qquad\qquad
   \times
   ( \, \Delta_{\bi}(\tau) - \bar{\Delta}_{\bi}(\tau) \, )
   ( \, \Delta_{\bj}(\tau') - \bar{\Delta}_{\bj}(\tau') \, )
   +
   \dotsb
   \notag
\end{align}
about the stationary points $\Delta = \bar{\Delta}$, defined by
\begin{equation}\label{BH.e:eq22}
   \Bigl [ \,
      \frac{ \delta S_{\text{eff}}[ \, \Delta;J,K \, ] }
           { \delta \Delta_{\bi}(\tau) } \,
   \Bigr ]_{\bar{\Delta}}
   =
   0 \>,   
\end{equation}
and computing the remaining path integral by the method of steepest descent, we find
\begin{align}
   &\beta \, \Omega[\, J,K \,]
   =
   S_{\text{eff}}[\, \bar{\Delta};J,K \, ]
   \label{BH.e:eq23} \\
   & \qquad
   +
   \frac{1}{2} \,
   \int_0^{\beta} \!\!\! \rd \tau \sum_{\bi} \,
   \Tr{ \Ln{ \calD^{-1}_{\bi,\bi}[\tau,\tau] } }
   +
   \dotsb \>,
   \notag \\
   &
   =
   -
   \frac{1}{2}
   \iint_0^{\beta} \!\!\! \rd \tau \, \rd \tau' \sum_{\bi,\bj} \>
   \bar{\Phi}_{\bi}^{\dagger}(\tau) \, 
   \calG_{\bi,\bj}^{-1}(\tau,\tau') \, 
   \bar{\Phi}_{\bj}(\tau')
   \notag \\
   & \quad
   -
   \int_0^{\beta} \!\!\! \rd \tau \sum_{\bi} \, 
   \Bigl \{ \,
      \frac{\bar{\chi}^2_{\bi}(\tau) - | \bar{A}_{\bi}(\tau)|^2}{2 U}
      +
      K^{\dagger}_{\bi}(\tau) \, \bar{\Delta}_{\bi}(\tau)
      \notag \\
      & \quad
      -
      \frac{1}{2} \,
      \Tr{ \Ln{ \calG_{\bi,\bi}^{-1}(\tau,\tau) } }
      -
      \frac{1}{2} \,
      \Tr{ \Ln{ \calD_{\bi,\bi}^{-1}(\tau,\tau) } } \,
   \Bigr \}
   \notag
\end{align}
where
\begin{equation}\label{BH.e:eq24}
   \calD^{-1}_{\bi,\bj}(\tau,\tau')
   =
   \Bigl [ \, 
      \frac{ \delta^2 S_{\text{eff}}[\, \Delta;J,K \, ] }
           { \delta \Delta_{\bi}(\tau) \, \delta \Delta_{\bj}(\tau') } \,
   \Bigr ]_{\bar{\Delta}} \>.
\end{equation}
Here $\bar{\Phi}_{\bi}(\tau)$ is defined as the solution of
\begin{equation}\label{BH.e:eq25}
   \bar{\Phi}_{\bi}(\tau)
   =
   \int_0^{\beta} \!\!\! \rd \tau' \sum_{\bj} \, \calG_{\bi,\bj}(\tau,\tau') \, J_{\bj}(\tau') \>,
\end{equation}
and is a functional of the currents $(J,K)$.  Explicitly, the stationary points are defined by the solutions of the equations
\begin{subequations}\label{BH.e:eq26}
\begin{align}
   \frac{\bar{\chi}_{\bi}(\tau)}{U}
   &=
   \frac{1}{2} \, \bar{\Phi}_{\bi}^{\dagger}(\tau) \, \bar{\Phi}_{\bi}(\tau) 
   +
   \frac{1}{2} \,
   \Tr{ \calG_{\bi,\bi}(\tau,\tau) } 
   +
   k_{0\,\bi}(\tau) \>,
   \label{BH.e:eq26a} \\
   \frac{\bar{A}_{\bi}(\tau)}{2 \, U}
   &=
   \frac{1}{2} \,
   \bar{\Phi}_{\bi}^{\dagger}(\tau) \, \sigma_{+} \bar{\Phi}_{\bi}(\tau) 
   +
   \frac{1}{2} \,
   \Tr{ \sigma_{-} \calG_{\bi,\bi}(\tau,\tau) } 
   -
   k_{\bi}(\tau) \>,
   \label{BH.e:eq26b}   
\end{align}
\end{subequations}
and are functionals of the currents and $\sigma_{\pm}$ are the Pauli matrices. Introducing the Legendre transformation,
\begin{align}
   \beta \, V_{\text{eff}}[ \, \Phi,\Delta \, ]
   &=
   \int_0^{\beta} \!\!\! \rd \tau \sum_{\bi} \, 
   \bigl [ \,
      J^{\dagger}_{\bi}(\tau) \, \Phi_{\bi}(\tau)
      +
      K^{\dagger}_{\bi}(\tau) \, \Delta_{\bi}(\tau) \,
   \bigr ]
   \notag \\
   & \qquad\qquad
   -
   \beta \, \Omega[\, J,K \,] \>,
   \label{BH.e:eq27}
\end{align}
we obtain the thermodynamic effective potential,
\begin{align}
   &V_{\text{eff}}[ \, \Phi,\Delta \, ]
   =
   \frac{1}{2 \beta}
   \iint_0^{\beta} \!\!\! \rd \tau \, \rd \tau' \sum_{\bi,\bj} \>
   \Phi_{\bi}^{\dagger}(\tau) \, 
   \calG_{\bi,\bj}^{-1}(\tau,\tau') \, 
   \Phi_{\bj}(\tau')
   \notag \\
   & \qquad
   +
   \frac{1}{\beta}
   \int_0^{\beta} \!\!\! \rd \tau \sum_{\bi} \, 
   \Bigl \{ \,
      \frac{\chi^2_{\bi}(\tau) - | A_{\bi}(\tau)|^2}{2 U}
      \label{BH.e:eq28} \\
      & \qquad\qquad\qquad\qquad
      +
      \frac{1}{2} \,
      \Tr{ \Ln{ \calG_{\bi,\bi}^{-1}(\tau,\tau) } }
   \Bigr \} \>,
   \notag
\end{align}
where we have dropped the trace-log term involved the $\calD$ propagator, which is higher order in our expansion.  The currents are now given by derivatives of $V_{\text{eff}}[ \, \Phi,\Delta \, ]$ with respect to the fields,
\begin{equation*}\label{BH.e:eq29}
   J^{\dagger}_{\bi}(\tau)
   =
   \beta N \, \frac{ \partial V_{\text{eff}}[ \, \Phi,\Delta \, ] }{\partial \Phi_{\bi}(\tau)} \>,
   \quad
   K^{\dagger}_{\bi}(\tau)
   =
   \beta N \, \frac{ \partial V_{\text{eff}}[ \, \Phi,\Delta \, ] }{\partial \Delta_{\bi}(\tau)} \>.
\end{equation*}
The thermodynamic potential is evaluated at zero currents, which is at the minimum of $V_{\text{eff}}[ \, \Phi,\Delta \, ]$.  The average particle number is given by
\begin{equation}\label{BH.e:eq30}
   N
   =
   -\frac{\partial V_{\text{eff}}[ \, \Phi,\Delta \, ]}{\partial \mu} 
\end{equation}
evaluated at the minimum of the effective potential.   

%
%
\section{\label{s:homosys}Homogeneous systems}

For homogeneous lattice systems in equilibrium, the fields are independent of $\tau$ and $\bi$.  Expanding the inverse Green function in a three dimensional Fourier series,
\begin{equation}\label{BH.e:eq31}
   \calG^{-1}_{\bi,\bj}(\tau,\tau')
   =
   \frac{1}{ \beta N_s^3}
   \sum_{\bk,n} \,
   \tilde{\calG}^{-1}_{\bk,n} \,
   e^{i [ \, 2\pi \bk \cdot ( \bi - \bj )/N_s - \omega_n \, ( \tau - \tau' ) \, ] } \>,
\end{equation}
where $\omega_n = 2 \pi n / \beta$ are the Bose Matsubara frequencies.  Here $\bk = (k_x,k_y,k_z)$ is a triplet of integers, each running from $-N_s/2$ to $N_s/2 - 1$.  The total number of sites in the cubic box is $N_s^3$ and the \emph{filling} factor, $\nu$, is defined to be the number of particles per site, $\nu = N/N_s^3$. 
From Eq.~\eqref{BH.e:eq6}, the Fourier transform of the Green function is given by
\begin{equation}\label{BH.e:eq33}
   \tilde{\calG}^{-1}_{\bk,n}
   = 
   \begin{pmatrix}
      \epsilon_{\bk} + \chi' - i \omega_n & -A \\[3pt]
      -A^{\ast} & \epsilon_{\bk} + \chi' + i \omega_n
   \end{pmatrix} \>,
\end{equation}
where we have put $\chi' = \sqrt{2} \, \chi - \mu$, and the kinetic energy is written in terms of the lattice momentum, $\hat \bk$, as
\begin{equation}\label{BH.e:eq34}
   \epsilon_{\bk}
   = J \, \hat{\bk}^2
   = 
   2 J \,
   \sum_{s=x,y,z} 
   [ \, 1 - \cos( 2 \pi \, k_s / N_s ) \, ]
   \>,
\end{equation}
Using standard techniques, 
\begin{align}
   &\frac{1}{2 \beta} \,
   \int_0^{\beta} \!\!\! \rd \tau \sum_{\bi} \, 
   \Tr{ \Ln{ \calG_{\bi,\bi}^{-1}(\tau,\tau) } }
   =
   \frac{1}{2 \beta}
   \sum_{\bk,n} \,
   \Tr{ \Ln{ \tilde{\calG}^{-1}_{\bk,n} } }
   \notag \\
   & \qquad
   =
   \frac{1}{2 \beta}
   \sum_{\bk,n} \,
   \Ln{ \Det{ \tilde{\calG}^{-1}_{\bk,n} } }
   =
   \frac{1}{2 \beta}
   \sum_{\bk,n} \,
   \Ln{ \omega_n^2 + \omega_k^2 }
   \notag \\
   & \qquad
   =
   \sum_{\bk} \,
   \Bigl \{ \,
      \frac{\omega_k}{2}
      +
      \frac{1}{\beta} \,
      \Ln{ 1 - e^{-\beta \omega_k} } \,
   \Bigr \} \>,
   \label{BH.e:eq37}
\end{align}
where
\begin{equation}\label{BH.e:eq38}
   \omega_k
   =
   \sqrt{ ( \, \epsilon_k + \chi' \, )^2 - | A |^2 } \>.
\end{equation}
The effective potential \eqref{BH.e:eq28} for the homogeneous case then becomes
\begin{align}
   &V_{\text{eff}}[ \, \Phi,\Delta \, ] / N_s^3
   \label{BH.e:eq39} \\
   & \quad
   =
   \chi' \, | \phi |^2
   -
   \frac{1}{2} \, [ \, A \, \phi^{\ast\,2} + A^{\ast} \phi^2 \, ]
   -
   \frac{ ( \, \chi' + \mu \, )^2 }{4 U}
   +
   \frac{ | A |^2 }{ 2 U }
   \notag \\
   & \quad
   +
   \frac{1}{N_s^3} 
   \sum_{\bk} \,
   \Bigl \{ \,
      \frac{1}{2} \,
      [ \,
         \omega_k
         -
         \epsilon_k
         -
         \chi' \,
      ]
      +
      \frac{1}{\beta} \,
      \Ln{ 1 - e^{-\beta \omega_k} } \,
   \Bigr \} \>.
   \notag
\end{align}
Here we have renormalized the effective potential by subtracting the zero-point energy.  The coupling constant is finite and does not need to be renormalized. 
Minimizing the effective potential with respect to the fields gives
\begin{subequations}\label{BH.e:eq40}
\begin{align}
   & ( \, \chi' - A^{\ast} \, ) \, \phi
   =
   0 \>,
   \label{BH.e:eq40a} \\
   & \frac{\chi' + \mu}{2 U}
   =
   | \phi |^2
   +
   \label{BH.e:eq40b} 
   \frac{1}{N_s^3} 
   \sum_{\bk} \,
   \Bigl \{ \,
      \frac{ \epsilon_k + \chi' }{ 2 \omega_k } \,
      [ \, 2 n_k + 1 \, ] 
      -
      \frac{1}{2} \,
   \Bigr \} \>,
   \\
   & \frac{A}{U}
   =
   \phi^2
   +
   \frac{A}{N_s^3} 
   \sum_{\bk} \,
   \frac{[ \, 2 n_k + 1 \, ]  }{ 2 \omega_k } \>,
   \label{BH.e:eq40c} 
\end{align}
\end{subequations}
where $n_k = 1 / [ \, e^{\beta \omega_k} - 1 \, ]$, and with the filling factor given by
\begin{equation}\label{BH.e:eq41}
   \nu
   =
   \frac{N}{N_s^3}
   =
   - \frac{1}{N_s^3} \, 
   \frac{\partial V_{\text{eff}}[ \, \Phi,\Delta \, ] }{ \partial \mu }
   =
   \frac{\chi' + \mu}{2 U} \>.
\end{equation}
Because of the $U(1)$ invariance of the Lagrangian, at the minimum of the potential we can choose $\phi$ to be real.  Then, from \eqref{BH.e:eq40a} $A$ is also real since $\chi'$ is real.  We interpret $| \phi |^2$ as the number of condensed particles per site, and put
\begin{equation}\label{ST.e:eq42}
   | \phi |^2 = \phi^2 = \nu_0 \, \frac{N_0}{N_s^3} = \nu \, n_0 \>,
\end{equation}
with the condensate fraction, $n_0 = N_0 /  N$.  The sums over $\bk$ then omit the $\bk = 0$ mode.  The gap equations then become
\begin{subequations}\label{BH.e:eq43}
\begin{align}
   \nu
   &=
   \nu n_0
   +
   \frac{1}{N_s^3}
   \sum_{\bk}{}' \,
   \Bigl \{ \,
      \frac{ \epsilon_k + \chi' }{ 2 \omega_k } \,
      [ \, 2 n_k + 1 \, ] 
      -
      \frac{1}{2} \,
   \Bigr \} \>,
   \label{BH.e:eq43a} \\
   \frac{A}{U}
   &=
   \nu n_0
   +
   \frac{A}{N_s^3}
   \sum_{\bk}{}' \,
   \frac{[ \, 2 n_k + 1 \, ]  }{ 2 \omega_k } \>,
   \label{BH.e:eq43b} 
\end{align}
\end{subequations}
where $\omega_k$ is given in Eq.~\eqref{BH.e:eq38} with $\epsilon_k$ given by Eq.~\eqref{BH.e:eq34}.  

For comparison purposes, we note that the LOAF effective potential per unit volume for the continuum case is: 
\begin{align}
   &V_{\text{eff}}[ \, \Phi,\Delta \, ] / V
   \label{cont.e:eq39} \\
   & 
   =
   \chi' \, | \phi |^2
   -
   \frac{1}{2} \, [ \, A \, \phi^{\ast\,2} + A^{\ast} \phi^2 \, ]
   -
   \frac{ ( \, \chi' + \mu \, )^2 }{4 \lambda}
   +
   \frac{ | A |^2 }{ 2 \lambda }
   \notag \\
   &
   + \!\!
   \int \!\! \frac{d^3 k}{(2\pi)^3} 
   \Bigl \{ 
      \frac{1}{2} 
      \Bigl [ 
         \omega_k
         -
         \epsilon_k
         -
         \chi' 
         +  \frac{ | A |^2 }{ 2 \lambda } 
      \Bigr ]
      +
      \frac{1}{\beta} 
      \Ln{ 1 - e^{-\beta \omega_k} } 
   \Bigr \} \>.
   \notag
\end{align}
which gives the equations 
\begin{subequations}\label{cont.e:eq40}
\begin{align}
   & ( \, \chi' - A^{\ast} \, ) \, \phi
   =
   0 \>, \quad
   \rho = \frac{\chi' + \mu}{2 \lambda} \>,
   \label{cont.e:eq40a} \\
   & \frac{\chi' + \mu}{2 \lambda}
   =
   | \phi |^2
   +
   \label{cont.e:eq40b} 
   \int \frac{d^3 k}{(2\pi)^3} \,
   \Bigl \{ \,
      \frac{ \epsilon_k + \chi' }{ 2 \omega_k } \,
      [ \, 2 n_k + 1 \, ] 
      -
      \frac{1}{2} \,
   \Bigr \} \>,
   \\
   & \frac{A}{\lambda}
   =
   \phi^2
   +
   \int \frac{d^3 k}{(2\pi)^3} \,
   \Bigl \{ \,
      \frac{1}{ 2 \omega_k } \,
      [ \, 2 n_k + 1 \, ] 
      -
      \frac{1}{2 \epsilon_k} \,
   \Bigr \} \>.
   \label{cont.e:eq40c} 
\end{align}
\end{subequations}
where the kinetic energy is the usual $\epsilon_k = \hbar^2 k^2/(2m)$. The differences between the two theories reduce to the naive substitution of the integral with the sum over the allowed momenta, an extra term in the renormalization of the effective potential, and the kinetic energy modification on the lattice. The continuum coupling constant, $\lambda=4\pi \hbar^2 \, a_0 / m$, corresponds to the Hubbard parameter, $U$, on the lattice. Here  $a_0$ is the $s$-wave scattering length in the dilute atomic Bose gas. The solutions II(i) and II(ii) correspond to first-order and second-order phase transitions, respectively.

%
%

\begin{figure}[t]
   \includegraphics[width=0.95\columnwidth]{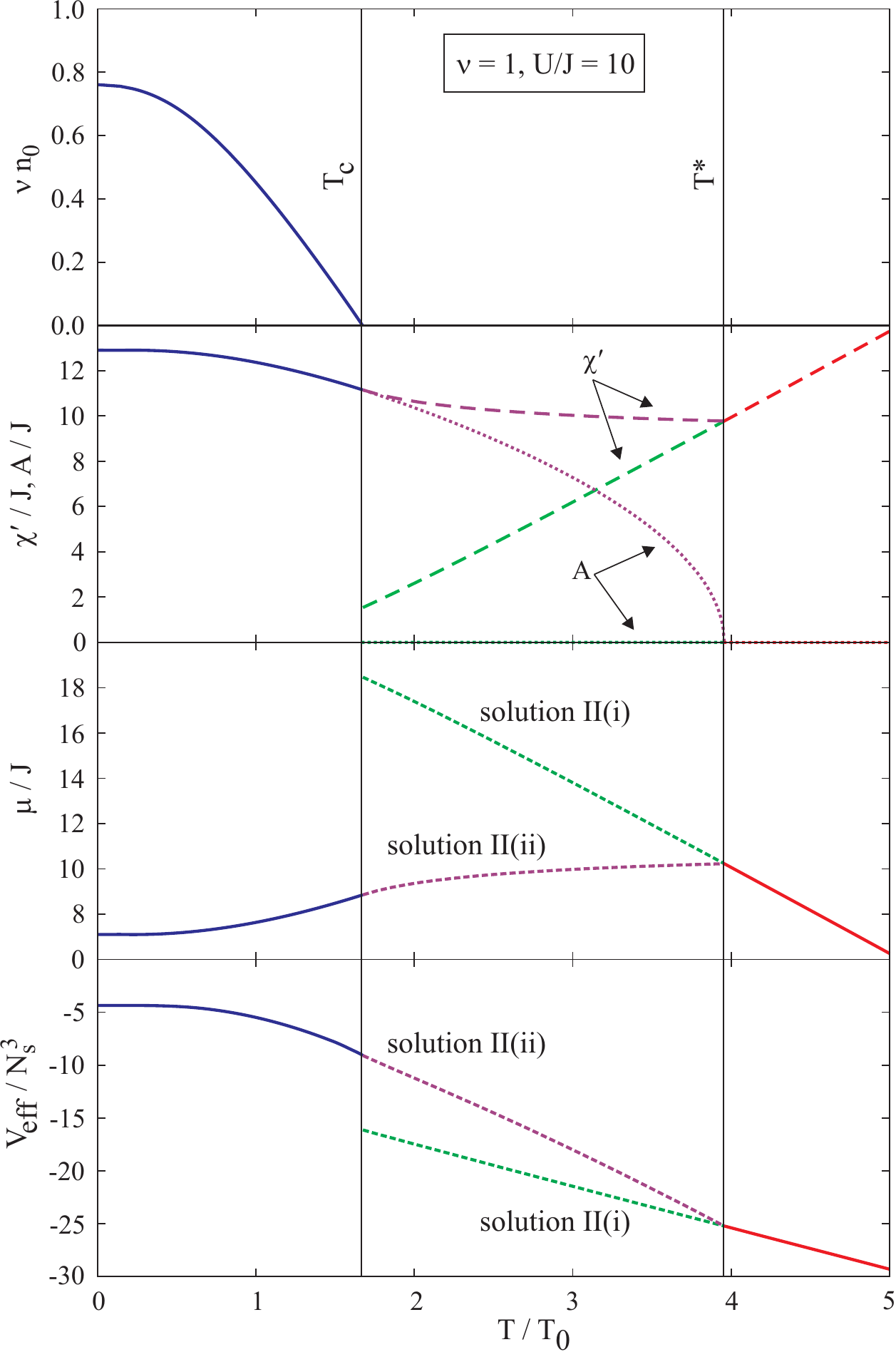}
   \caption{\label{fig1}(Color online) 
   Temperature dependence of the condensate fraction, $\nu n_0$, scaled auxiliary fields, $\chi'$ and $A$, scaled chemical potential, $\mu$, and effective potential, $V_{\text{eff}}$, for a Hubbard interaction parameter value, $U/J=10$, at unity filling, $\nu=1$. Here, the temperature, $T$, is scaled by $T_0$, the critical temperature of the ideal (non-interacting) Bose-Hubbard model. Solutions~II(i) and II(ii), corresponds to first- and second-oreder phase transitions, respectively.
    }
\end{figure} 

\begin{figure}[t]
   \includegraphics[width=0.95\columnwidth]{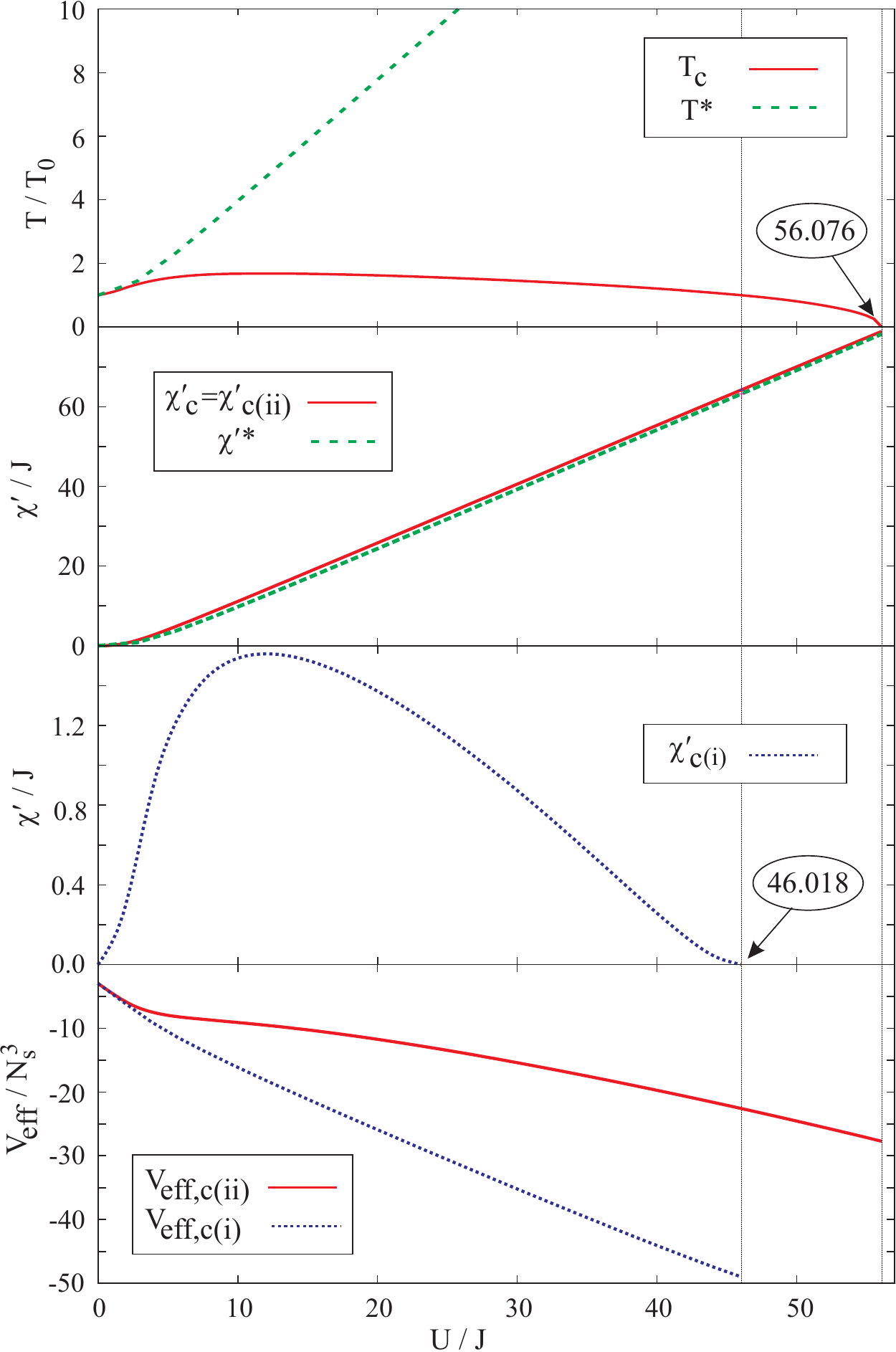}
   \caption{\label{fig2}(Color online) 
   LOAF scaled critical temperatures, $T_c$ and $T^\star$, and the corresponding scaled normal-density auxiliary fields, $\chi'_c=\chi'_{c(ii)}$, $\chi'{}^\star$, and $\chi'_{c(i)}$, and effective potentials, as a function of the Bose-Hubbard model coupling constant, ($U/J$), at unity filling, $\nu=1$. We note that only the critical temperature, $T_c$, features a downturn with the interaction strength and leads to a quantum phase transition (QPT) for $(U/J)_c \approxeq 56.07$ where $T_c$ goes to zero. In addition, the normal-density auxiliary field $\chi'_{c(i)}$ for solution~II(i) is not defined for $T_c < T_0$, where $T_0 \approxeq 5.59$ is the critical temperature of the non-interacting Bose system. Therefore, LOAF predicts a critical point (CP) at coordinates $T_\text{CP} = T_0$ and $(U/J)_\text{CP} = 46.02$. For coupling values $(U/J)_\text{CP} < (U/J) \leq (U/J)_c$, the transition is second-order and LOAF predicts a \emph{diatom} condensate $A \neq 0$ in the absence of the \emph{usual} Bose-Einstein condensate fraction for temperatures $T_c < T < T_0$.  
Because $T^\star > T_0$, the system is in the normal phase for $T > T_0$ in this region. 
For coupling constants $(U/J) < (U/J)_\text{CP}$ the transition is first-order because $V_\text{eff,c(i)} < V_\text{eff,c(ii)}$ in that region.  
    }
\end{figure} 

\begin{figure}[t]
   \includegraphics[width=0.95\columnwidth]{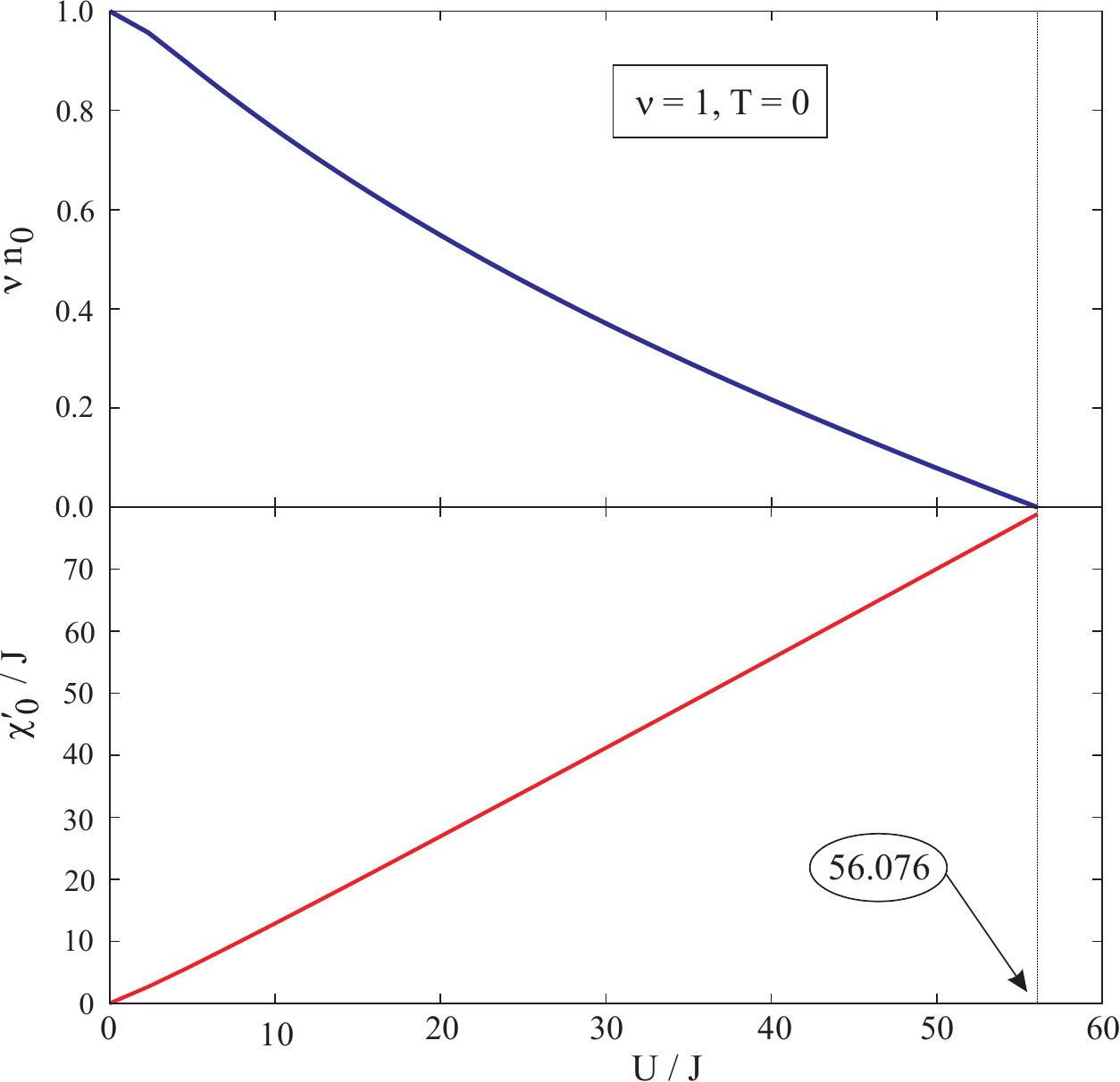}
   \caption{\label{fig0}(Color online) 
   Interaction strength dependence of the zero-temperature condensate fraction, $\nu n_0$, and the corresponding scaled normal-density auxiliary field, $\chi'_0$, at unity filling, $\nu=1$.
    }
\end{figure} 

\begin{figure}[t]
   \includegraphics[width=0.95\columnwidth]{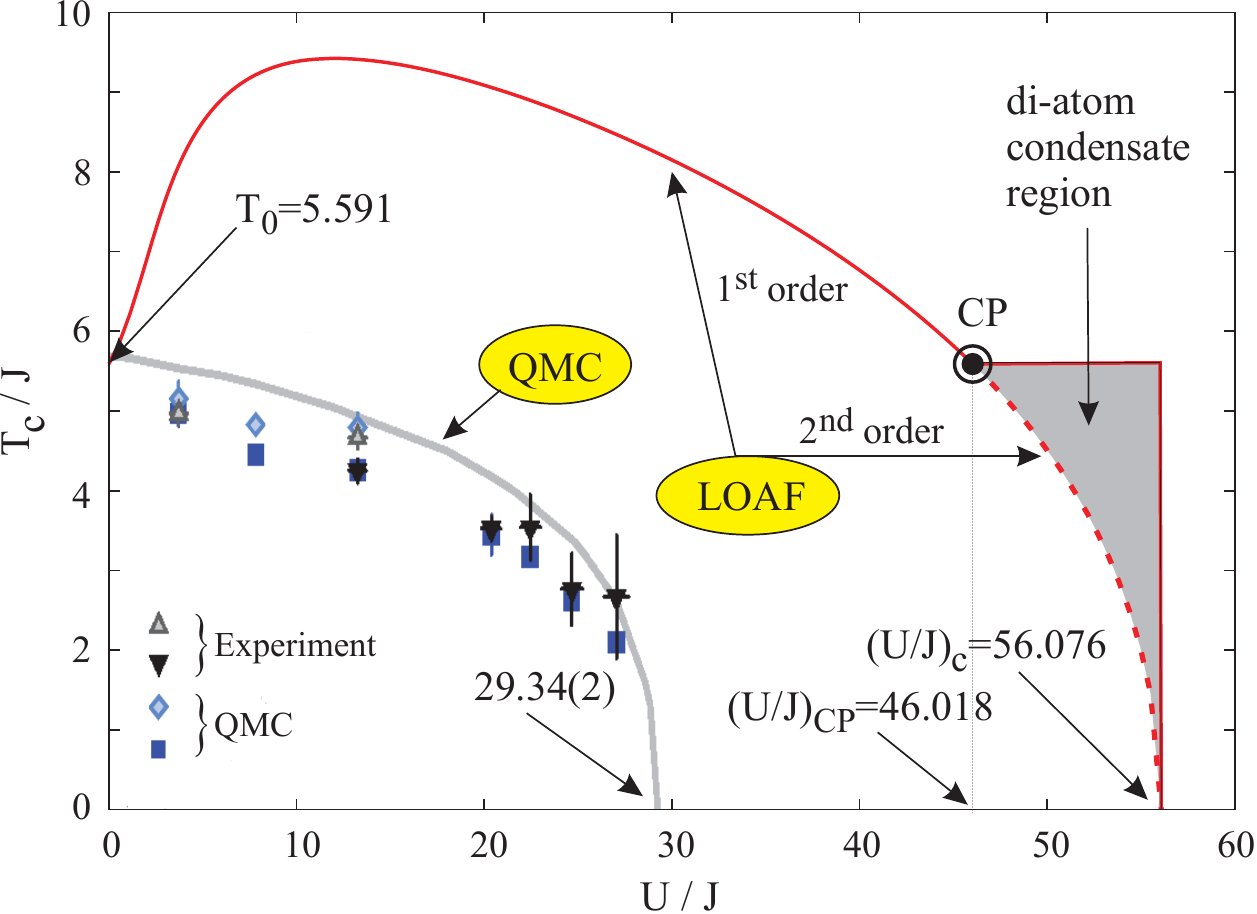}
   \caption{\label{fig3}(Color online) 
   Comparison of the coupling constant dependence of the LOAF critical temperature, $T_c$, at unity filling, $\nu=1$, with experimental~\cite{r:Trotzky:2010fk}  and quantum Monte Carlo (QMC) results~\cite{PhysRevB.75.134302}. The LOAF value of the critical Hubbard parameter value, $(U/J)_c=56.076$, should be compared to the QMC critical value, $(U/J)_c=29.34(2)$, reported in Ref.~\onlinecite{PhysRevB.75.134302}. LOAF also predicts a critical point at $(U/J)_\text{CP} = 46.02$. The solid and dashed lines indicate first- and second-order phase transitions predicted by LOAF theory, respectively. The shaded area depicts the region where a \emph{diatom} condensate without the \emph{usual} Bose-Einstein condensate is expected.
    }
\end{figure} 

\begin{figure}[t!]
   \includegraphics[width=0.95\columnwidth]{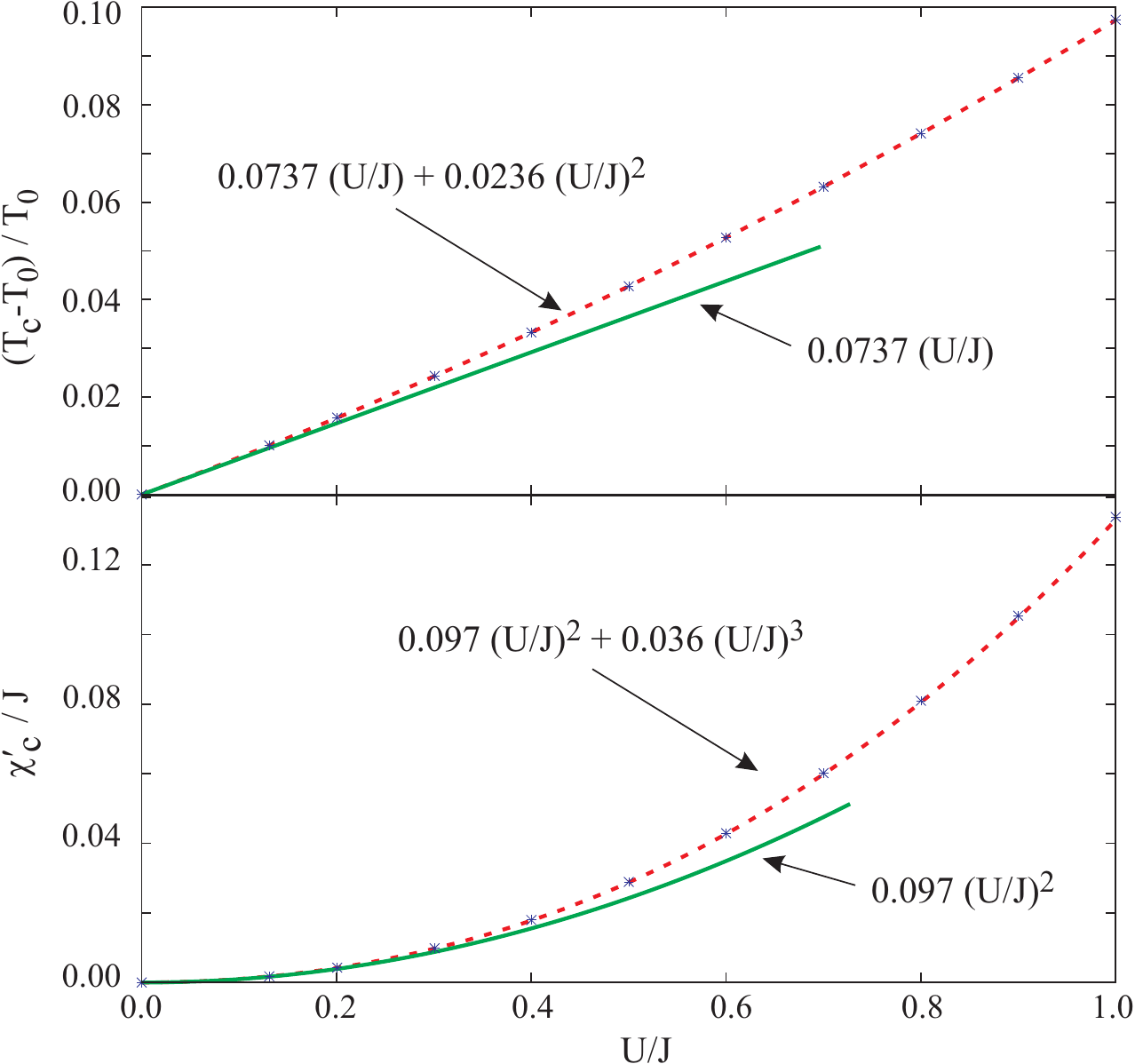}
   \caption{\label{fig4}(Color online) 
   Weak-coupling limit behavior of the critical temperature, $T_c$, and corresponding critical value of the scaled normal-density auxiliary field, $\chi'_c$, as a function of the Hubbard interaction parameter value, $U/J$, at unity filling, $\nu=1$.
    }
\end{figure} 

\begin{figure}[t]
   \includegraphics[width=0.95\columnwidth]{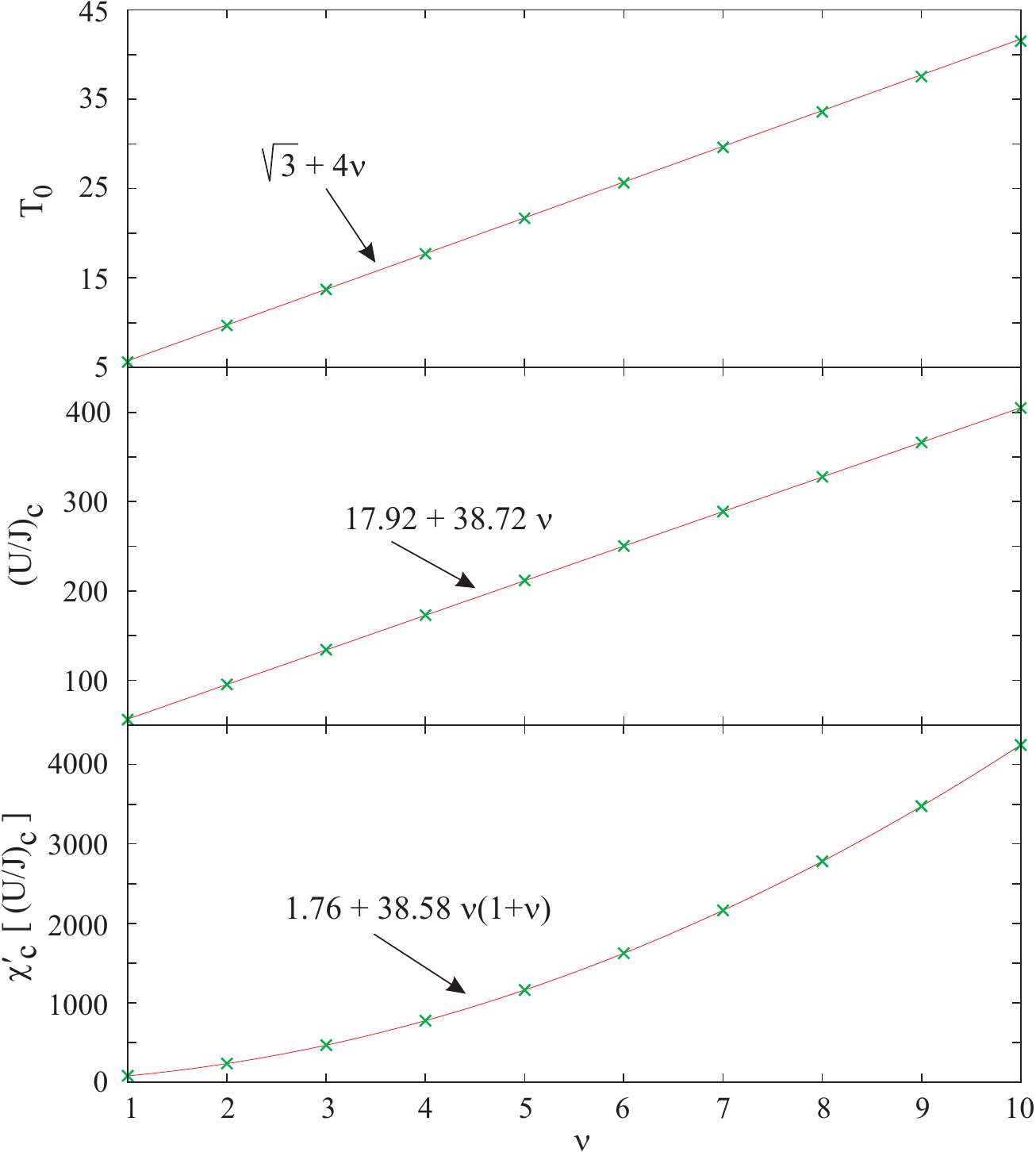}
   \caption{\label{fig5}(Color online) 
   Ideal gas critical temperature, $T_0$, critical value of the scaled Hubbard parameter, $(U/J)_c$, and the corresponding critical value of the scaled normal-density auxiliary field, $\chi_c'$, as a function of the filling factor,~$\nu$.
    }
\end{figure}

%
%
\section{\label{s:numerics}Results and Discussions}

The numerical analysis of the solutions space for Eqs.~\eqref{BH.e:eq43}, leads to three distinct regions in the Bose-Hubbard model phase diagram:
\begin{enumerate}
\item[I.] The broken symmetry case where $\phi \ne 0$ and $\chi' = A$.  Then $\omega = \sqrt{ \epsilon_k ( \epsilon_k + 2 \chi' )}$.  In this region, we solve the equations~\footnote{Eqs.~\eqref{BH.e:eq44} are identical with Eqs.~(62) in Ref.~\onlinecite{r:Kleinert:2013fk}, where Kleinert \emph{et al.}  considered only this region in developing their Hamiltonian ``two-collective'' field theory of the BH model.}:
\begin{subequations}\label{BH.e:eq44}
\begin{align}
   \nu
   &=
   \nu n_0
   +
   \frac{1}{N_s^3}
   \sum_{\bk}{}' \,
   \Bigl \{ \,
      \frac{ \epsilon_k + \chi' }{ 2 \omega_k } \,
      [ \, 2 n_k + 1 \, ] 
      -
      \frac{1}{2} \,
   \Bigr \} \>,
   \label{BH.e:eq44a} \\
   \frac{\chi'}{U}
   &=
   \nu n_0
   +
   \frac{\chi'}{N_s^3}
   \sum_{\bk}{}' \,
   \frac{[ \, 2 n_k + 1 \, ]  }{ 2 \omega_k } \>.
   \label{BH.e:eq44b} 
\end{align}
\end{subequations}
\item[II.] The case when $\phi = 0$ so that $n_0 = 0$, and either
\begin{enumerate}
\item[(i)] $A=0$ so that $\omega_k = \epsilon_k + \chi'$ and 
\begin{equation}\label{BH.e:eq45}
   \nu
   =
   \frac{1}{N_s^3}
   \sum_{\bk}{}' \
   n_k \>.
\end{equation}
This solution corresponds to a first-order phase transition. Eq.~\eqref{BH.e:eq45} does not depend on the interaction strength and applies for temperatures, $T \ge T_c$, where $T_c$ is the critical temperature defined by the zero condensate faction limit, $n_0 \rightarrow 0$, in Eqs.~\eqref{BH.e:eq44}.
\item[(ii)] or $0 \le A \le \chi'$ so that $\omega_k = \sqrt{ ( \epsilon_k + \chi' )^2 - A^2 }$, and 
\begin{subequations}\label{BH.e:eq46}
\begin{align}
   \nu
   &=
   \frac{1}{N_s^3}
   \sum_{\bk}{}' \,
   \Bigl \{ \,
      \frac{ \epsilon_k + \chi' }{ 2 \omega_k } \,
      [ \, 2 n_k + 1 \, ] 
      -
      \frac{1}{2} \,
   \Bigr \} \>,
   \label{BH.e:eq46a} \\
   \frac{1}{U}
   &=
   \frac{1}{N_s^3}
   \sum_{\bk}{}' \,
   \frac{1}{ 2 \omega_k } \, [ \, 2 n_k + 1 \, ]
   \>.
   \label{BH.e:eq46b}    
\end{align}
\end{subequations}
This solution corresponds to a second-order phase transition.
\end{enumerate}
\item[III.] the normal case where $\phi = 0$ and $A = 0$.  In this case we solve Eq.~\eqref{BH.e:eq45} as in case II(i) above.
\end{enumerate}
We note that the LOAF solutions for the cubic lattice are identical with the LOAF solutions for the continuum system~\cite{PhysRevLett.105.240402,PhysRevA.83.053622,PhysRevA.85.023631}.
  
To make contact with Ref~\onlinecite{r:Kleinert:2013fk}, we convert the finite sums over~$\bk$ to integrals by defining $\bq = 2 \bk / N_s$, so that formally in Eqs.~\eqref{BH.e:eq43} we substitute 
\begin{equation}\label{BH.e:eq47}
   \frac{1}{N_s^3} \sum_{\bk}
   \Rightarrow
   \iiint_{-1}^{+1} \frac{\rd^3 q}{8}
   =
   \iiint_{0}^{+1} \!\!\! \rd^3 q \>.
\end{equation}
This substitution is exact in the limit $N_s \rightarrow \infty$.
In the following, all quantities other than the filling factor, $\nu$, and the condensate fraction, $n_0$, can be scaled by $J$ without loss of generality.

We define the critical temperature $T_c$ as the point in region~I where the \emph{usual} condensate fraction, $n_0$, vanishes.  The second critical temperature, $T^{\star}$,  is the temperature where the \emph{diatom} condensate, $A$, vanishes in region~II(ii).  In the non-interacting limit, $U \rightarrow 0$, we have $T^\star \rightarrow T_c$.

The critical temperature $T_c$ and fields $\chi'_c=A_c$ are given by the solution of Eqs.~\eqref{BH.e:eq45} in region~I when $n_0 = 0$:
\begin{subequations}\label{BH.e:eq50}
\begin{align}
   \nu
   &=
   \iiint_{0}^{+1} \!\!\! \rd^3 q \,
   \Bigl \{ \,
      \frac{ \epsilon_q + \chi_c' }{ 2 \omega_q } \,
      [ \, 2 n_q + 1 \, ] 
      -
      \frac{1}{2} \,
   \Bigr \} \>,
   \label{BH.e:eq50a} \\
   \frac{1}{U}
   &=
   \iiint_{0}^{+1} \!\!\! \rd^3 q \,
   \frac{[ \, 2 n_q + 1 \, ]  }{ 2 \omega_q } \>,
   \quad
   n_q
   =
   \frac{1}{e^{\omega_q / T_c} - 1} \>,
   \label{BH.e:eq50b} 
\end{align}
\end{subequations}
with $\omega_q = \sqrt{\epsilon_q ( \epsilon_q + 2 \chi'_c)}~$\footnote{Eqs.~\eqref{BH.e:eq50} should be contrasted with Eqs.~(97) and~(98) in Ref.~\onlinecite{r:Kleinert:2013fk}. The latter are inconsistent in the limit $T_c \rightarrow 0$, because the right side of those equations vanish in that limit. Nonetheless, the critical value for $\nu=1$ of the Hubbard parameter reported in Ref.~\onlinecite{r:Kleinert:2013fk} was 28.04.}. 
From Eqs.~\eqref{BH.e:eq50}, the critical value of the Hubbard parameter, $(U/J)_c$, is obtained by taking the limit $T_c \rightarrow 0$. For $\nu=1$ we obtain the critical Hubbard parameter value $(U/J)_c \approxeq 56.076$, to be compared with the critical value of 29.34(2) obtained by \emph{ab initio} quantum Monte Carlo simulations~\cite{PhysRevB.75.134302}.

The critical temperature $T^{\star}$ and field $\chi'{}^\star$ are defined by the solution of Eqs.~\eqref{BH.e:eq46} when $n_0 = 0$ and $A = 0$:
\begin{subequations}\label{BH.e:eq50.1}
\begin{gather}
   \nu
   =
   \iiint_{0}^{+1} \!\!\! \rd^3 q \, n_q \>,
   \qquad
   n_q
   =
   \frac{1}{e^{\omega_q / T^{\star}} - 1} \>,
   \label{BH.e:eq50.1a} \\
   \frac{1}{U}
   =
   \iiint_{0}^{+1} \!\!\! \rd^3 q \,
   \frac{[ \, 2 n_q + 1 \, ]  }{ 2 \omega_q } \>,
   \label{BH.e:eq50.1b} 
\end{gather}
\end{subequations}
with $\omega_q = \epsilon_q + \chi'{}^\star$.

In Fig.~\ref{fig1} we illustrate the temperature dependence of the condensate fraction, $\nu n_0$, scaled auxiliary fields, $\chi'$ and $A$, scaled chemical potential, $\mu$, and effective potential, $V_{\text{eff}}$, for a Hubbard interaction parameter value, $U/J=10$, at unity filling, $\nu=1$. We find that in region~II, we have two possible solutions of the LOAF equations, as discussed above. The solution~II(i) gives rise to discontinuities in the temperature dependence of the auxiliary fields and chemical potential that lead to a discontinuity in the effective potential as well. This behavior is characteristic to a first-order phase transition. In contrast, the solution~II(ii) corresponds to a second-order phase transition, because the temperature dependence of $\chi'$, $A$, $\mu$, and $V_\text{eff}$ is smooth across $T_c$.  

For convenience, we denote by $\chi'_{c\text{(i)}}$ the value of the normal-density auxiliary field corresponding to the first-order phase transition solution~II(i) (see Eq.~\eqref{BH.e:eq45}) for $T=T_c$, and we introduce the notation $\chi'_{c\text{(ii)}}=\chi'_c$ to indicate the value of $\chi'$ at $T_c$ corresponding to the second-order phase transition solution~II(ii). We have, $\chi'_{c\text{(i)}} \rightarrow \chi'_{c\text{(ii)}}$ in the non-interacting limit, $U \rightarrow 0$. We recall that Eq.~\eqref{BH.e:eq45} is independent of the Hubbard parameter $U$ and is restricted to temperatures $T \ge T_c$. In the non-interacting limit we have $T_c \rightarrow T_0$, where $T_0$ is the critical temperature of the non-interacting lattice Bose system, and the non-interacting limit corresponds to $\chi' \rightarrow 0$. So, we find that $\chi'_{c\text{(i)}} \rightarrow 0$ in the limit $\{ U \rightarrow 0,~T_c \rightarrow T_0 \}$. Therefore a first-order phase transition may occur at $T_c$ only for an interaction strength, $U$, that gives a critical temperature $T_c \ge T_0$. Furthermore, the solution~II(i) is only possible for a temperature $T \ge T_0$. We will use this important observation next.

The coupling constant dependence of the LOAF critical temperatures, $T_c$ and $T^\star$, the corresponding scaled normal-density auxiliary fields, $\chi'_{c\text{(i)}}$, $\chi'_{c\text{(ii)}}$=$\chi'_c$, and $\chi'{}^\star$, and the corresponding effective potentials at $T_c$ are depicted in Fig.~\ref{fig2}. We find that the critical temperature~$T^\star$ increases monotonically with the BH model coupling constant, ($U/J$), whereas the critical temperature, $T_c$, increases with the interaction strength for $U/J \lesssim 10$ and then decreases with $(U/J)$. At unity filling, $\nu=1$, the critical temperature, $T_c$, goes to zero, for a critical value, $(U/J)_c \approxeq 56.076$. It is important to note that in the continuum case of a homogenous system of ultracold Bose atomic gases neither of the two critical temperatures $T_c$ and $T^\star$ goes to zero in LOAF~\cite{PhysRevA.84.023603}. It appears that in LOAF the presence of a quantum phase transition is related to the reduction in the allowed momentum-vector phase space. The latter is a consequence of the bosonic atoms being spatially confined on the lattice.

For completeness in Fig.~\ref{fig0} we illustrate the scaled Hubbard parameter, $U/J$, dependence of the zero-temperature condensate fraction, $\nu n_0$, and the corresponding scaled normal-density auxiliary field, $\chi'_0$, at unity filling, $\nu=1$.

As discussed above, the normal-density auxiliary field $\chi'_{c(i)}$ for the first-order phase transition solution~II(i) is not defined for $T_c < T_0$. That leads to the possibility of a critical point (CP) at coordinates $T_\text{CP} = T_0$ and $(U/J)_\text{CP} \approxeq 46.02$.  

Recalling that in LOAF the superfluid density is proportional to the square of the anomalous-density auxiliary field,~$A$ (see discussion in Ref.~\onlinecite{PhysRevA.86.013603}), it follows that the phase diagram of the Bose-Hubbard model in LOAF features two regions: 
First, for coupling values $(U/J) < (U/J)_\text{CP}$,  both solutions II(i) and II(ii) are possible and we may have either a first-order or a second-order phase transition solution. Because $V_\text{eff,c(i)} < V_\text{eff,c(ii)}$, LOAF predicts a first-order phase transition from the superfluid to the normal phase in this region. This scenario corresponds to the solution II(i). 
Second, for coupling values $(U/J)_\text{CP} < (U/J) < (U/J)_c$, we have $T_c < T_0$ and the solution~II(i) is not possible for temperatures $T_c < T < T_0$. Hence, the transition is second-order as described by solution~II(ii). Because $T^\star > T_0$ for all couplings, solution~II(ii) applies only for temperatures $T_c < T < T_0$. In this temperature range LOAF predicts a diatom condensate $A \neq 0$ in the absence of the \emph{usual} Bose-Einstein condensate fraction and the system is in a superfluid state for all temperatures $0 < T < T_0$.  As seen in Fig.~\ref{fig1}, for all temperatures $T_c < T < T^\star$ we have $V_\text{eff,(i)} < V_\text{eff,(ii)}$. Therefore at~$T_0$ the system undergoes a first-order phase transition from the superfluid to the normal phase.

The LOAF phase diagram for the BH model can be compared with predictions of quantum Monte Carlo (QMC) simulations~\cite{PhysRevB.75.134302}. The depression in the critical temperature was observed experimentally in ultracold Bose atom systems in three-dimensional lattices~\cite{r:Trotzky:2010fk} and  the QMC compares well with experiments for couplings $(U/J) \lesssim 20$. The QPT predicted by Monte Carlo occurs for  $(U/J)_c=29.34(2)$, so about half the critical value predicted by LOAF. In Fig.~\ref{fig3} we show that our LOAF results compare qualitatively well with existing experimental and \emph{ab initio} QMC results~\cite{r:Trotzky:2010fk,PhysRevB.75.134302}. The shaded area in Fig.~\ref{fig3} is the region where a \emph{diatom} condensate is expected to be present in the system in the absence of the \emph{usual} Bose-Einstein condensate~\footnote{This is contrary to the conclusions of Kleinert \emph{et al.}~\cite{r:Kleinert:2013fk}. Those authors did not consider the theory for temperatures $T>T_c$.}. 

Just like in the continuum case, the LOAF results for the lattice show that the critical temperature, $T_c$, departs from the ideal gas result (see Fig.~1). Numerical results depicted in Fig.~\ref{fig4} show that in the weak-coupling limit $(\Delta T_c)$ increases linearly with the coupling $(U/J)$ with a slope parameter $\approxeq 0.0737$. Furthermore, we find that the critical value of the auxiliary field, $\chi_c$, is proportional with the square of the coupling, $(U/J)^2$. 

Finally, in Fig.~\ref{fig5} we plot the ideal gas critical temperature, $T_0$, critical value of the scaled Hubbard parameter, $(U/J)_c$, and the corresponding critical value of the scaled normal-density auxiliary field, $\chi_c'$, as a function of the filling factor,~$\nu$.

%
%
\section{\label{s:conclusions}Conclusions}

In this paper we developed the leading-order auxiliary field approximation (LOAF) for the Bose-Hubbard model corresponding to a system of Bose atoms confined in a three-dimensional lattice. The auxiliary-field formalism treats on equal footing condensates associated with the normal and anomalous densities.

For temperatures $T<T_c$ we showed that LOAF is the same as the ``two-collective'' field theory introduced by Kleinert \emph{et al.}~\cite{r:Kleinert:2013fk}. Here $T_c$ is the temperature where the \emph{usual} Bose-Einstein condensate vanishes. For temperatures $T>T_c$, LOAF has two possible solutions corresponding to either a first-order or a second-order phase transition. 
For both solutions, the superfluid state is indicated by the presence of an anomalous-density \emph{diatom} condensate in the system. 

The BH phase diagram in LOAF features a line of first-order transitions ending in a critical point at $T_\text{CP} = T_0$ and a finite coupling $(U/J)_\text{CP}$. In the first-order phase transition solution, the diatom condensate auxiliary field $A$ vanishes at $T_c$, so the system evolves from a superfluid to a normal phase. First-order phase transition solutions are limited to temperatures $T > T_0$, where $T_0$ is the critical temperature of the non-interacting system.

Beyond the critical point, the transition is second order. In the case of the second-order phase transition, the diatom condensate auxiliary field, $A$, goes to zero smoothly and vanishes at the critical temperature, $T^\star$. For $T_c < T < T^\star$, the system is still in the superfluid phase, because in LOAF the superfluid density is proportional to the square of $A$, and not to the \emph{usual} condensate fraction, $n_0$. For $T>T^\star$, we have $A=0$ and the system is in the normal phase. This scenario provides for the second-order phase transition known to take place in liquid helium and dilute gases of Bose atoms.  The critical temperature $T^\star$ does not vanish either in the continuum or the lattice cases.
In the case of the BH model, $T^\star$ is always greater than $T_0$, so the system never reaches $T^\star$, but rather exhibits a first-order phase transition at $T_0$ from the superfluid to the normal phase. Contrary to the conclusions of Kleinert \emph{et al.}~\cite{r:Kleinert:2013fk}, for couplings $(U/J)_\text{CP} < (U/J) < (U/J)_c$ and temperatures $T_c < T < T_0$ we have a region where a \emph{diatom} condensate is expected in the absence of the \emph{usual} Bose-Einstein condensate.

For Bose systems on a lattice the critical temperature $T_c$ goes to zero for a finite value of the Hubbard interaction parameter, $(U/J)_c$, indicating a quantum phase transition (QPT) similar to the superfluid to Mott insulator transition. For continuum systems, $T_c$ does not vanish~\cite{PhysRevA.84.023603} and LOAF predicts no QPT in the case of infinite Bose matter. So in LOAF the QPT is due to the spatial confinement of the Bose atoms on the lattice. The LOAF phase diagram of the BH model compares qualitatively well with existing experimental and \emph{ab initio} quantum Monte Carolo results~\cite{r:Trotzky:2010fk,PhysRevB.75.134302}. 

It is clear that in order to understand the LOAF phase diagram of strongly interacting systems of particles it is necessary to extend this non-perturbative theory beyond the mean-field level of approximation discussed so far.

%
%

\begin{acknowledgments}
This work was performed in part under the auspices of the National Science Foundation and the US Department of Energy. 
\end{acknowledgments}

%
%
\appendix
%
%
\section{\label{s:noninteracting}Non-interacting case}

The non-interacting case, $U/J = 0$, is recovered by setting $\chi \rightarrow 0$ and $A \rightarrow 0$ in Eq.~\eqref{BH.e:eq39}.  Then $\chi' = - \mu$, $\omega_k \rightarrow \epsilon_k - \mu$, and the effective potential becomes
\begin{align}
    V_{\text{eff}}[ \, z,T \, ]
    &=
    \frac{1}{\beta}
    \sum_{\bk}
    \Ln{ 1 - z e^{ -\beta \epsilon_k } }
    \label{ST.e:NI.eq1} \\
    &=
    \frac{1}{\beta} \, \Ln{1-z}
    +
    \frac{1}{\beta}
    \sum_{\bk}{}'
    \Ln{ 1 - z e^{ -\beta \epsilon_k } } \>,
    \notag
\end{align}
where the primed sum omits the $\bk = 0$ term and the fugacity $z$ is defined by
\begin{equation}\label{ST.e:NI.eq2}
   z
   =
   e^{\beta \mu} \>,
   \qquad
   0 \le z \le 1 \>.
\end{equation}
The particle number is
\begin{align}
   N
   &=
   - \frac{\partial V_{\text{eff}}[ \, z,T \, ]}{\partial \mu}
   =
   - \beta z \, \frac{\partial V_{\text{eff}}[ \, z,T \, ]}{\partial z}
   \label{ST.e:NI.eq3} \\
   &=
   N_0
   +
   \sum_{\bk}{}' \,
   \frac{z}{e^{\beta \epsilon_k} - z} \>,
   \notag
\end{align}
with $N_0 = z / ( 1 - z )$.  Dividing by $N_s^3$ and replacing the sums over $\bk$ by an integral over $\rd^3 q$ gives
\begin{subequations}\label{ST.e:NI.eq4}
\begin{align}
   \nu
   &=
   \nu n_0
   +
   F(z,T) \>,
   \label{ST.e:NI.eq4a} \\
   F(z,T)
   &=
   \frac{1}{N_s^3}
   \sum_{\bk}{}' \,
   \frac{z}{e^{\beta \epsilon_k} - z}
   \label{ST.e:NI.eq4b} \rightarrow
   \iiint_{0}^{+1} \!\!\! \rd^3 q \,
   \frac{z}{e^{\beta \epsilon_q/T} - z}
   \notag
\end{align}
\end{subequations}
The critical point is where $z \rightarrow 1$, in which case
\begin{equation}\label{ST.BH.e:eq4}
   \nu
   =
   \nu n_0
   +
   F(1,T) 
\end{equation}
is an equation giving the condensate fraction $n_0$ as a function of $T$.  The maximum value of $T = T_0$ is when $n_0 = 0$, and is the solution of the equation
\begin{equation}\label{ST.BH.e:eq5}
   \nu
   =
   F(1,T_0) 
\end{equation}
for fixed value of $\nu$.  
For $\nu = 1$, we have $T_0/J = 5.591$ as expected~\cite{PhysRevB.75.134302}.

%
%
%
\bibliography{johns.bib}
%
%
\end{document}

%% file: Johnsdefs.tex
\usepackage[T1]{fontenc}    
\usepackage{amsfonts}
\usepackage{amssymb}
\usepackage{amsmath}
\usepackage{amsthm}
\usepackage{amscd}
\usepackage{color}
\usepackage{graphicx}
\usepackage{epsfig}
\usepackage{subfigure}
\usepackage{fancyhdr}
\usepackage{dcolumn}
\usepackage{time}
\usepackage{hyperref}
\hypersetup{
    unicode=false,          
    pdftoolbar=true,        
    pdfmenubar=true,        
    pdffitwindow=false,     
    pdfstartview={FitH},    
    pdftitle={Paper},       
    pdfauthor={John Dawson},     
    pdfsubject={Phys Rev article},   
    pdfcreator={John Dawson},   
    pdfproducer={John Dawson}, 
    pdfkeywords={keyword1} {key2} {key3}, 
    pdfnewwindow=true,      
    colorlinks=true,        
    linkcolor=red,          
    citecolor=blue,         
    filecolor=magenta,      
    urlcolor=cyan           
}
\definecolor{dark}{gray}{.5}
\definecolor{light}{gray}{.75}
\definecolor{darkmagenta}{rgb}{.5,0,.5}
\newcolumntype{d}[1]{D{.}{.}{#1}}
\newcommand{\QED}%
   {$\mathcal{Q\kern-.1em \lower.6ex\hbox{$\mathcal{E}$}\kern-.1667em D}$}
\newcommand{\QCD}%
   {$\mathcal{Q\kern-.1em \lower.6ex\hbox{$\mathcal{C}$}\kern-.1667em D}$}

\newcommand{\SUSYext}%
   {$\mathcal{S \kern-0.08em \lower 0.5ex \hbox{$\mathcal{U}$}
   \kern-0.05em S \kern-0.2em \lower 0.5ex
   \hbox{$\mathcal{Y}$}}\kern-0.05em{}_{\text{ext}}$}
\theoremstyle{definition}

\theoremstyle{remark}

\newcommand{\rD}{\text{D}}

\newcommand{\bi}{\mathbf{i}}
\newcommand{\bj}{\mathbf{j}}
\newcommand{\bk}{\mathbf{k}}

\newcommand{\bq}{\mathbf{q}}

\newcommand{\bx}{\mathbf{x}}

\newcommand{\calD}{\mathcal{D}}

\newcommand{\calG}{\mathcal{G}}

\newcommand{\rd}{\mathrm{d}}
\newcommand{\Partial}[4]
   {\Bigl ( \frac{\partial #1 }{\partial #2 } \Bigr )_{\! #3, #4 }}
\newcommand{\Det}[1]{\det [ \, #1 \, ]}

\newcommand{\Ln}[1]{\ln [ \, #1 \, ]}

\newcommand{\Tr}[1]{\mathrm{Tr} [ \, #1 \, ]}

\newcommand{\bra}[1]%
   {\ensuremath{\langle \, #1 \, |}}
\newcommand{\Bra}[1]%
   {\ensuremath{\langle \, #1 \, |}}
\newcommand{\bigbra}[1]%
   {\ensuremath{\Bigl \langle \, #1 \, \Bigr |}}
\newcommand{\ket}[1]%
   {\ensuremath{| \, #1 \, \rangle}}
\newcommand{\Ket}[1]%
   {\ensuremath{| \, #1 \, \rangle}}
\newcommand{\bigket}[1]%
   {\ensuremath{\Bigl | \, #1 \, \Bigr \rangle}}
\newcommand{\braket}[2]%
   {\ensuremath{\langle \, #1 \, | \, #2 \, \rangle}}
\newcommand{\Braket}[2]%
   {\ensuremath{\langle \, #1 \, | \, #2 \, \rangle}}
\newcommand{\matrixelement}[3]%
   {\ensuremath{\langle \, #1 \, | \, #2 \, | \, #3 \, \rangle}}
\newcommand{\MEangle}[3]%
   {\ensuremath{\langle \, #1 \, | \, #2 \, | \, #3 \, \rangle}}
\newcommand{\MatEl}[3]%
   {\ensuremath{\langle \, #1 \, | \, #2 \, | \, #3 \, \rangle}}
\newcommand{\MEparen}[3]%
   {\ensuremath{( \, #1 \, | \, #2 \, | \, #3 \, )}}
\newcommand{\pbra}[1]%
   {\ensuremath{( \, #1 \, |}}
\newcommand{\pket}[1]%
   {\ensuremath{| \, #1 \, )}}    
\newcommand{\Brap}[1]%
   {\ensuremath{( \, #1 \, |}}
\newcommand{\Ketp}[1]%
   {\ensuremath{| \, #1 \, )}}    
\newcommand{\pbraket}[2]%
   {\ensuremath{( \, #1 \, | \, #2 \, )}}
\newcommand{\braV}[1]%
   {\ensuremath{\langle \, #1 \, \Vert}}
\newcommand{\ketV}[1]%
   {\ensuremath{\Vert \, #1 \, \rangle}}
\newcommand{\Comm}[2]%
   {\ensuremath{[ \, #1, #2 \, ]}}
\newcommand{\AntiComm}[2]%
   {\ensuremath{\{ \, #1, #2 \, \}}}
\newcommand{\Pbracket}[2]%
   {\ensuremath{\{ \, #1, #2 \, \} }}
\newcommand{\PBracket}[2]%
   { \ensuremath{ \{ \, #1, #2 \, \}_{\lower1.0ex\hbox{\scriptsize \text{PB}}} } }
\newcommand{\wedgeComm}[2]
   {\ensuremath{[ \, #1, #2 \, ]_{\lower1.0ex\hbox{\scriptsize $\wedge$}} }}
\newcommand{\Expect}[1]
   {\ensuremath{\langle \, #1 \,  \rangle}}
\newcommand{\expect}[1]%
   {\ensuremath{\langle \, #1 \,  \rangle}}
\newcommand{\Expectbig}[1]%
   {\ensuremath{\Bigl \langle \, #1 \, \Bigr \rangle}}
\newcommand{\expectbig}[1]%
   {\ensuremath{\Bigl \langle \, #1 \, \Bigr \rangle}}
\newcommand{\expectc}[2]%
   {\ensuremath{\langle \, \{ \, #1 , #2 \, \} \, \rangle}}
\newcommand{\expectq}[2]%
   {\ensuremath{\langle \, [ \, #1 , #2 \, ] \, \rangle}}
\newcommand{\expectT}[1]%
   {\ensuremath{\langle \, \mathcal{T} \{ \, #1 \, \} \, \rangle}}
\newcommand{\expectaT}[1]%
   {\ensuremath{\langle \, \mathcal{T}^{\ast} \{ \, #1 \, \} \, \rangle}}
\newcommand{\expectTbig}[1]%
   {\ensuremath{\biggl \langle \, \mathcal{T}  \biggl \{ \, #1 \,%
\biggr \} \, \biggr \rangle}}
\newcommand{\expectTabig}[1]%
   {\ensuremath{\biggl \langle \, \mathcal{T}^{\ast} \biggl \{ \, #1 \,%
\biggr \} \, \biggr \rangle}}
\newcommand{\Tproduct}[1]%
   {\ensuremath{\mathcal{T} \{ \, #1 \, \} } }
\newcommand{\aTproduct}[1]%
   {\ensuremath{\mathcal{T}^{\ast} \{ \, #1 \, \} } }
\newcommand{\Nproduct}[1]%
   {\ensuremath{\mathcal{N} \{ \, #1 \, \} } }
\newcommand{\ctpTproduct}[1]%
   {\ensuremath{\mathcal{T}_{\mathcal{C}} \{ \, #1 \, \} }}
\newcommand{\tauordered}[1]%
   {\ensuremath{\mathcal{T}_{\tau} \{ \, #1 \, \} }}
\newcommand{\Norder}[1]%
   {\ensuremath{: #1 : \, }}
\newcommand{\expectTproduct}[1]%
   {\ensuremath{\langle \, \mathcal{T} \{ \, #1 \, \} \, \rangle}}
\newcommand{\expectTCproduct}[1]%
   {\ensuremath{\langle \, \mathcal{T}_{\mathcal{C} \{ \, #1 \, \} \, \rangle}}}
\newcommand{\expectComm}[2]%
   {\ensuremath{\langle \, [ \, #1 , #2 \, ] \, \rangle}}
\newcommand{\expectPbracket}[2]%
   {\ensuremath{\langle \, \{ \, #1 , #2 \, \} \, \rangle}}
\newcommand{\threej}[6]%
{\begin{pmatrix} #1 & #2 & #3 \\ #4 & #5 & #6 \end{pmatrix}}
\newcommand{\sixj}[6]%
{\begin{Bmatrix} #1 & #2 & #3 \\ #4 & #5 & #6 \end{Bmatrix}}
\newcommand{\ninej}[9]%
{\begin{Bmatrix} #1 & #2 & #3 \\ #4 & #5 & #6 \\%
 #7 & #8 & #9 \end{Bmatrix}}
\newcommand{\reducedme}[3]%
{\langle \, #1 \, \Vert \, #2 \, \Vert \, #3 \, \rangle }